\documentclass{article}

\usepackage{graphicx}
\usepackage{fullpage,amsmath,amssymb,latexsym}
\usepackage{multirow}
\usepackage{natbib}

\begin{document}

\title{Constraining Saturn's Core Properties by a Measurement of Its Moment of Inertia - Implications to the Cassini Solstice Mission}


\author{R. Helled$^1$\\
Department of Earth and Space Sciences\\
University of California, Los Angeles, CA 90095 1567, USA\\
rhelled@ucla.edu}

\date{}
\maketitle 

\begin{abstract}
Knowledge of Saturn's axial moment of inertia can provide valuable information on its internal structure. We suggest that Saturn's angular momentum may be determined by the Solstice Mission (Cassini XXM) by measuring Saturn's pole precession rate and the Lense-Thirring acceleration on the spacecraft, and therefore, put constraints on Saturn's  moment of inertia. 
It is shown that Saturn's moment of inertia can change up to $\sim$2\% due to different core properties. However, a determination of Saturn's rotation rate is required to constrain its axial moment of inertia. A change of about seven minutes in rotation period leads to a similar uncertainty in the moment of inertia value as different core properties (mass, radius). A determination of Saturn's angular momentum and rotation period by the Solstice Mission could reveal important information on Saturn's internal structure, in particular, its core properties. 
\end{abstract}


\section{Introduction}
The Cassini spacecraft has provided enormous improvement of our understanding of Saturn, its rings, and satellites. In the context of interior modeling, Cassini has provided accurate measurements of Saturn's gravitational coefficients $J_2$, $J_4$ and $J_6$ that are used to constrain the planet's internal structure and composition. The Cassini extended extended mission (XXM), also known as the Solstice Mission, is planed to have 42 close orbits that would enable a determination of Saturn's gravitational harmonics up to degree 10. In addition, the final close orbits to Saturn might also provide constraints on Saturn's internal rotation period by detecting small perturbations from Saturn's axisymmetric magnetic field (Spilker, 2010). \par

It has been suggested that the Juno spacecraft (Bolton, 2006) could provide a measurement of Jupiter's axial moment of inertia by including the effects of pole precession and Lense-Thirring acceleration in the Doppler signal analysis (Helled et al., 2011). This is because both of these effects include the planet's angular momentum, and therefore its normalized moment of inertia, NMoI=$C/MR^2$ where $C$ is the axial moment of inertia, and $M$ and $R$ are the planet's mass and mean radius, respectively. Since the final orbits of the Solstice mission are essentially similar in nature to the orbits of Juno around Jupiter, it is certainly feasible to measure Saturn's angular momentum. \par

Like Jupiter, Saturn's axis is tilted to its orbital plane, and the torque exerted from the Sun on Saturn and its equatorial satellites results in a precession of Saturn's axis of rotation. Since Saturn's orbital plane is not fixed in space, the precession rate of Saturn's pole is not constant, but is changing periodically, slowly over time (Ward \& Hamilton, 2004). Since the pole precession rate is dependent on the planet's NMoI, an accurate determination of Saturn's pole location can constrain Saturn's pole precession rate, and therefore its moment of inertia. In addition to pole precession, information on Saturn's moment of inertia can be obtained by measuring the spacecraft acceleration due to the Lense-Thirring effect. The Lense-Thirring effect is a general relativistic effect that describes the amount of coordinate frame dragging by a rotating body (Lense \& Thirring, 1918). A point mass (the Cassini spacecraft) gains Lense-Thirring acceleration due to its close orbits to the massive rotating Saturn (e.g., Iorio, 2010). As a result, including the Lense-Thirring effect and Saturn's pole precession in the Doppler signal analysis can lead to a determination of Saturn's NMoI.\par

It should be noted, however, that the quantity that is actually measured by the spacecraft is the planet's spin angular momentum $L=C\cdot \omega=$NMoI$\cdot MR^2 \cdot \omega$ where $\omega$ is the rotation rate. The value of NMoI cannot be measured separately from $L$. Saturn's rotation period is not well constrained within a few minutes, and this uncertainty introduces an uncertainty in the NMoI value. It is shown below that if indeed Saturn's rotation period is constrained by magnetic field measurements during the Solstice Mission, tighter constraints on Saturn's NMoI can be introduced.
\par

The measured planetary gravitational coefficients $J_{2n}$ are typically used to constrain the planet's internal structure (e.g., Guillot, 2005), in addition, the high-order harmonics can be used to constrain the depth of differential rotation within the planet (Hubbard, 1982, 1999). A measurement of NMoI could provide an additional constraint on Saturn's interior structure, in particular, its core properties (Helled et al., 2011). This is due to the fact that there is {\it no} one-to-one correspondence between the gravitational harmonics and the moment of inertia as suggested by the Radau-Darwin relation (Radua, 1885; Darwin, 1900; Zharkov and Trubitsyn, 1978). Although the gravitational harmonics are strongly related to NMoI, more than one moment of inertia value can fit to a given gravitational field, and as a result, the NMoI introduces an additional constraint for interior models (Helled et al., 2011). Below we investigate how Saturn's NMoI value changes with different core properties (mass, radius) for two different solid-body rotation periods, and explore whether a measurement of Saturn's NMoI can provide information on its core properties. \par

\section{Saturn's Moment of Inertia}
We use a simple core/envelope model in which the internal density profile is divided to a constant density core, and an envelope for which the density is represented by a 6th order polynomial. We make sure that the density is deceasing  monotonically towards the planetary surface. For a given core mass and radius, the polynomial coefficients are iterated until Saturn's internal density reproduces Saturn's measured gravitational coefficients.  The interior model uses Saturn's mean radius in the traditional manner of the theory of figures level-surfaces (Zharkov and Trubitsyn, 1978). Once a density profile that matches Cassini's measured gravitational field is found, the normalized axial moment of inertia is computed. \par

We fit Saturn's measured $J_2$ and $J_4$ {\it exactly}, and $J_6$ within its measured error bar. This approach allows us to test the variation of the NMoI value for Saturn's gravitational field as measured by Cassini. Saturn's measured $J_{2n}$ are given in Jacobson et al.~(2006) and in units of $10^{-6}$ for a reference equatorial radius of 60,330 km are $J_2=16290.71\pm0.27$, $J_4=-935.8\pm2.8$, $J_6=86.1\pm9.6$. We restrict our models to core mass 0 M$_{\oplus}$ $<$M$_C$ $<$ 30 M$_{\oplus}$ and core density $\rho_C$ $<$ 30 g cm$^{-2}$. We vary Saturn's core density and core radius and search for density profiles that satisfy the constraints listed above. More details on the model can be found in Helled et al. (2011) and references therein. \par

Since Saturn's rotation period is unconstrained within a few minutes (e.g., Gurnett et al., 2007), and the interior model uses the planet's rotation rate (and mean radius) we use two different solid-body rotation periods. The first is Voyager's rotation period of 10h 39m 24s (Saturn's system III rotation period). The mean radius corresponding to that period is found to be 58182.96 km (Helled et al., 2009a, 2009b). The second period is about seven minutes shorter, and is taken to be 10h 32m 35s, hereafter AS rotation period (Anderson and Schubert, 2007). The shorter rotation period suggested by Anderson and Schubert (2007) was found to minimize Saturn's dynamical heights, and to produce a wind profile similar to Jupiter in which winds go both east and west, with more moderate wind velocities (Read et al., 2009). The mean radius for AS rotation period is 58255.4 km. Since different rotation periods lead to different shapes, and therefore different equatorial radii (see Helled et al., 2009b for details), the values of $J_{2n}$ (which are given for an arbitrary reference equatorial radius) change as well. Table 1. summarizes the physical parameters that are used by the interior model for the two rotation periods. 
We consider the two solid-body rotation periods to demonstrate how a change of a few minutes in Saturn's rotation period affects its NMoI. It is possible, however, that the Voyager and AS rotation periods do not actually span the possible range of interior rotation periods. In fact, it is also possible that Saturn does not rotate uniformly as a solid-body but instead rotates on cylinders to a certain depth. If the planet rotates differentially on cylinders the rotation rate must be represented by a function instead of a constant, and corrections to the interior model must be applied (Hubbard, 1999). In order to fully constrain Saturn's NMoI value a better determination of its rotation period/profile is required.\par

\par 

\subsection{Results} 
The interior model results for the two solid-body rotation periods we consider are presented in Figure 1. The top figure presents all the allowed models in the core mass - core radius - NMoI parameter space. 
It can be seen from the top figure that the range of NMoI values for a given rotation period changes up to $\sim$ 2\%. Saturn's NMoI values range from 0.215 - 0.220, and 0.213 - 0.218 for the Voyager and AS rotation periods, respectively. Saturn's NMoI is defined as C/MR$^2$, where $R$ being the mean radius, and is different for the two periods. This difference, however, leads to a shift of only about 0.02\% in the NMoI value, ten times smaller than the variation in NMoI value for different core parameters. While typically lower NMoI values correspond to massive cores, this is not always the case. A given NMoI value can correspond to a large range of core masses and radii. While a given NMoI value cannot perfectly constrain Saturn's internal structure, the results suggest that an accurate determination of Saturn's axial moment of inertia could still provide valuable constraints on its core properties. Also suggested by the figure is that lower NMoI values typically  correspond to shorter rotation periods (Helled et al., 2008). However, a change of about seven minutes in rotation period leads to only a few percents change in NMoI. \par

The bottom figure shows the allowed models in the core mass - core radius - angular momentum ($L$) parameter space. It is clear from the figure that there is a significant overlap between the model results for the two rotation periods. An uncertainty of about seven minutes in rotation period is essentially comparable to the uncertainty in Saturn's NMoI value due to different core properties. As a result, the range of the derived values of Saturn's angular momentum ($L=$NMoI$\cdot MR^2\cdot\omega$) is the same for the two cases despite a few narrow regions. The bottom figure demonstrates the degeneracy in determining Saturn's NMoI value by angular momentum measurements when the rotation period is unknown within a few minutes. Only when Saturn's rotation period is determined accurately by an independent method (e.g., magnetic field measurements) can one use angular momentum measurements to constrain Saturn's internal structure. 

\section{Summary}
Cassini may be able to determine Saturn's angular momentum, and therefore put constraints on its axial moment of inertia during the last close orbits around Saturn during the XXM (Solstice Mission). Such a measurement could provide important constraints on Saturn's internal structure, although the uncertainty in Saturn's rotation period leads to an uncertainty in its NMoI value, and therefore an independent measurement of Saturn's rotation period is desirable.  
Our model suggests that varying Saturn's core properties (mass, radius) results in a change of the NMoI value of the order of 2\%. A change of about seven minutes in solid-body rotation period leads to a shift of additional $\sim$ 2\% in NMoI, with a decrease in the NMoI value for the shorter rotation period.\par 

Constraining Saturn's NMoI value by a measurement of its angular momentum is possible only if the angular momentum is measured to a few tenths of percent, and an accurate determination of Saturn's rotation period is available.  A better determination of Saturn's rotation period also has important implications on Saturn's atmosphere dynamics (Anderson and Schubert, 2007; Read et al., 2009). Knowledge of Saturn's NMoI value can also be used to put constraints on dynamical models (Boue et al., 2009 and references therein) and to better understand and constrain planetary migration in the early solar system, put limits on the migration timescale, and on Neptune's primordial inclination (Boue et al., 2009). It is certainly desirable to preform measurements to determine Saturn's NMoI independently of its gravitational field, if possible. \par

\subsection*{Acknowledgments}
I acknowledge support from NASA through the Southwest Research Institute. 
I thank an anonymous referee for valuable comments and suggestions, and John D. Anderson for fruitful discussions. 
 

\newpage

\begin{table}[h!]
\begin{center}
{\renewcommand{\arraystretch}{0.6}
\vskip 8pt
\begin{tabular}{l c c}
Rotation Period & 10h 32m 35s & 10h 39m 24s
\\[1pt]
\hline
\\[1pt]
$GM$ (km$^3$ s$^{-2}$) & 37931208. & 37931208. \\
\\[2pt]
$R$ (km) & 58165.6 & 58094.2\\
\\[2pt]
$a$ (km) & 60256.9 & 60141.4 \\
\\[2pt]
$J_2$ (10$^{-6}$) & 16330.2 & 16393.1 \\
\\[2pt]
$J_4$ (10$^{-6}$) &-940.4& -947.6 \\
\\[2pt]
$J_6$ (10$^{-6}$) & 86.77 & 87.77 \\
\hline
\end{tabular} 
}
\end{center}
\caption{\label{geoid} 
Physical parameters used by the interior model. The radii correspond to the 1 bar pressure-level. The harmonic coefficients $J_{2n}$ are obtained from the measured values which are given for a reference equatorial radius of 60,330 km (Jacobson et al.~2006), according to $a^{2n}J_{2n}$=(60,330 km)$^{2n}J_{2n}$(measured). Since the equatorial radius $a$ of the reference geoid depends on the rotation period, so do the values of $J_{2n}$.     
} 
\end{table}

\newpage
\begin{figure}
   \centering
    \includegraphics[width=5.in]{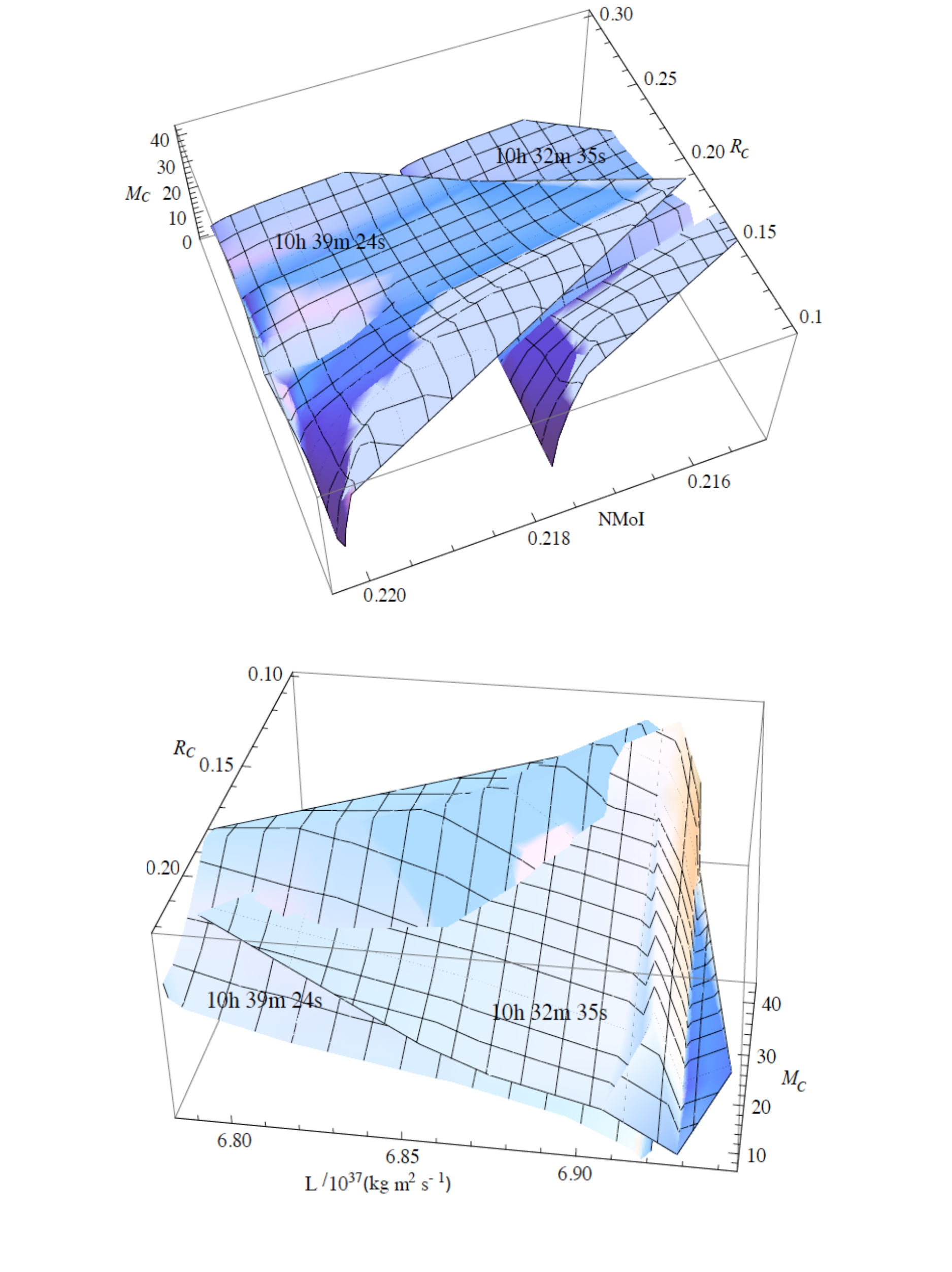}
    \caption[sat]{top: Saturn's allowed models for solid-body rotation periods 10h 39m 24s (Voyager) and 10h 32m 35s (AS).  Shown are the normalized core radius ($R_C$), core mass ($M_C$) in Earth masses and the normalized axial moment of inertia NMoI. bottom: Saturn's allowed models for the two rotation periods. The normalized core radius ($R_C$), core mass ($M_C$) in Earth masses, and Saturn's angular momentum $L$ (kg m$^2$ s$^{-1}$) are presented. All the derived models fit Saturn's gravitational field, mass, and shape, and in addition have core properties within the allowed bounds (i.e., core mass $<$ 30 M$_{\oplus}$; core density $<$ 30 g cm$^{-3}$).}
\end{figure}

\end{document}